# DESIGN AND TEST OF 704 MHz AND 2.1 GHz NORMAL CONDUCTING CAVITIES FOR LOW ENERGY RHIC ELECTRON COOLER

Binping Xiao[†], S. Belomestnykh[*], J. M. Brennan, J. C. Brutus, G. McIntyre, K. Mernick, C. Pai, K. Smith, T. Xin, A. Zaltsman

*Brookhaven National Laboratory, Upton, New York 11973-5000, USA*

V. Veshcherevich

*Cornell University, Ithaca, New York 14853, USA*

The Low Energy RHIC electron Cooler (LEReC) is currently under commissioning at BNL to improve RHIC luminosity for heavy ion beam energies below 10 GeV/nucleon. The linac of LEReC consists of a DC photoemission gun, one 704 MHz superconducting radio frequency (SRF) booster cavity, and three normal conducting cavities. It is designed to deliver a 1.6 MeV to 2.6 MeV electron beam, with peak-to-peak momentum spread $dp/p$ of less than $7 \cdot 10^{-4}$. Two of the three normal conducting cavities will be used in LEReC for energy spread correction. A single-cell 704 MHz cavity for energy de-chirping and a three-cell 2.1 GHz third harmonic cavity for RF curvature correction. In this paper, we present the designs and RF test results of these two cavities.

## I. Introduction

A bunched beam electron cooler called LEReC [1] has been designed and is under construction at the Relativistic Heavy Ion Collider (RHIC) to significantly improve the collider luminosity at energies below 10 GeV/nucleon to map the QCD phase diagram, especially to search the QCD critical point.

The electron linac of LEReC consists of a DC photoemission gun, one 704 MHz SRF booster cavity [2], and three normal conducting cavities: a 2.1 GHz third harmonic cavity, a 704 MHz cavity [3], and a 9 MHz cavity. The 704 MHz SRF booster cavity accelerates 400 keV bunches from the DC gun near the top of the RF wave, with an accelerating voltage up to 2.2 MV. The booster also introduces an energy chirp for ballistic stretching of the bunches in a beam transport section. The remaining curvature of the bunch in longitudinal phase space is then corrected by decelerating the beam in the 2.1 GHz third harmonic cavity, so that the bunch is nearly linear in phase space. After the 30-meter long transport section, the bunch is still short enough so that the energy chirp can be removed with the normal conducting 704 MHz cavity.[4] This linac is designed to operate in four modes: three pulsed modes with a 40% duty factor at 1.6, 2.0 and 2.6 MeV, and a CW electron beam mode at 1.6 MeV [1]. The tunable 9 MHz ferrite cavity [5] is designed for compensation of beam energy modulation due to transient beam loading of the SRF booster cavity in pulsed modes. Neither the SRF booster cavity nor the 9 MHz cavity is presented in this paper, which is dedicated to 2.1 GHz and 704 MHz normal conducting cavities.

The paper is organized as following: first, the RF design of the cavities is introduced; then, the energy spread caused by Higher Order Modes (HOMs), as well as possible Same Order Modes (SOMs, $\pi/3$ and $2\pi/3$ $TM_{010}$ modes that have the same order as the $TM_{010}$ $\pi$ mode in a 3-cell cavity), is considered and is fed back to the design of the cavities; after that, the frequency tuner and the Fundamental Power Coupler (FPC) designs are presented; thermal and mechanical simulation are described in Section VI, followed by the RF test results in the last section before the Conclusions section. The RF design optimization was performed using CST Microwave Studio® [6], and the final designs were simulated using ACE3P package [7]. Wake field simulation was done using CST Particle Studio® [6]. Multipacting was simulated using ACE3P package, and then cross-checked with a GPU-based code [8]. Thermal and mechanical simulations were performed using ANSYS™ [9].

## II. RF design of the cavities

### A. 2.1 GHz Cavity

The 2.1 GHz cavity is a three-cell cavity operating at 2.112 GHz. It delivers an accelerating voltage up to 250 kV, at 9 kW power dissipation in the cavity walls with an unloaded quality factor $Q$ of 20,000, and 21.2 kW maximum power from the beam (decelerating). The 704 MHz cavity is a single-cell structure operating at 704.0 MHz that delivers a voltage up to 251 kV, at 9.5 kW power dissipation in the cavity walls with an unloaded $Q$ of 34,000, and 13.4 kW maximum power to the beam (accelerating). Parameters of these two cavities are listed in Table 1.

[†]binping@bnl.gov
[*]present address: Fermi National Accelerator Laboratory, Batavia, IL 60510-5011, USA

The cross section view of the 2.1 GHz cavity is shown in Figure 1. A pillbox shape cell is adopted in this design. The wall between the adjacent cells is 10 mm thick, with the length of the vacuum portion of the center cell of ($\lambda/2 - 10$) mm, where $\lambda$ is the wavelength. Cell-to-cell coupling is determined by a 47.6 mm diameter drift tube between the cells. Cells with different nose cone heights, shown in Figure 1(I), were simulated to improve the cavity shunt impedance $R_0$ at $f_0 = 2.112$ GHz. For each simulation, the cavity radius is adjusted so that the $TM_{010}$ $\pi$-mode resonance frequency is maintained at $f_0$. The simulations showed maximum $R_0$ at a 2.5 mm nose cone height. For the end cells, the nose cone close to the center cell side is chosen to be the same as that for the center cell. No nose cone on the beam pipe side of the end cells is designed to simplify the cavity construction without significant degradation of $R_0$. The length of the end cell is chosen to be ($\lambda/2 - 10$) mm, and the beam pipe on each end cell is 48.9 mm in diameter, to suppress the bunch-to-bunch (inter-bunch) energy spread, as shown in the next section.

Table 1. Parameters of the 2.1 GHz and 704 MHz normal conducting cavities.

| Cavity | 2.1 GHz | 704 MHz |
|---|---|---|
| Frequency [MHz] | 2,112.0 | 704.0 |
| Tuning range [MHz] | -1.2 to +2.9 | -1.0 to +1.7 |
| Number of cells | 3 | 1 |
| Voltage [kV] | 250 | 251 |
| R/Q [Ohm] | 350 | 192 |
| $Q_0$ | 20,000 | 34,000 |
| FPC $Q_{ext}$ | 20,000 | 14,000 |
| Cavity power loss [kW] | 9.0 | 9.5 |
| Max beam power [kW] | -21.2 | 13.4 |
| RF amplifier power [kW] | 14 | 65 |

Two RF pickup ports, one is shown in Figure 1(J), are designed on the center cell of the cavity. A hook-shaped coupler is used to get a 10 W RF power out of the cavity at 250 kV. This 10 W power will get significantly attenuated due to losses in an RF cable from the cavity location at the RHIC Interaction Region 2 (IR2) to the low-level RF (LLRF) control board outside the RHIC tunnel. The output power can be adjusted by rotating the hook coupler.

With the 48.9 mm diameter beam pipe, it is easy to evacuate the cavity directly from the beam pipe. Thus, there is no pumping port on the cavity body.

An RF window (Figure 1(B), with details shown in Figure 8 (left)) with a transition taper to WR430 waveguide is attached to the cavity via a 90-degree, 50.1 mm radius E-bend waveguide. A 15.9 mm blending radius is applied all around the RF coupling slot located on the center cell. A V-shaped FPC port is shown in Figure 1(G), with a zoom-in view in Figure 8 (right). Details of the FPC design are described in Section V.

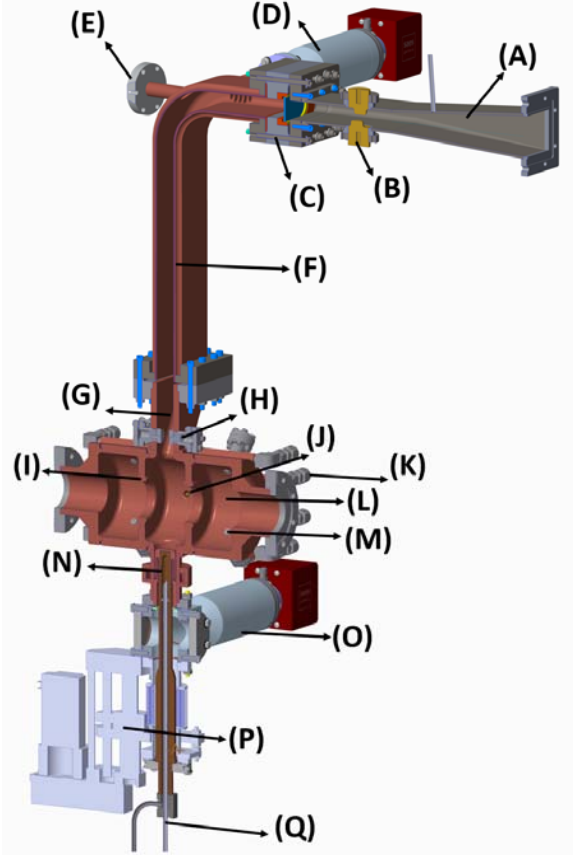

Figure 1. 2.1 GHz cavity cross section view. (A) waveguide adaptor from JLab530 to WR430; (B) bolt with a non-concentric knob; (C) JLab C100 RF window; (D) vacuum pump near the RF window; (E) view port; (F) FPC waveguide; (G) FPC port; (H) FPC tuner; (I) nose cone; (J) pickup coupler; (K) cavity water cooling channel; (L) cavity body; (M) fixed tuner; (N) main frequency tuner; (O) vacuum pump at the main tuner; (P) driver for the main tuner; (Q) main tuner water cooling channel.

### B. 704 MHz Cavity

The 704 MHz normal conducting cavity serves not only as a de-chirping cavity, but also an accelerating cavity. Depending on the kinetic energy of the electron beam, the voltage and phase requirements change. The requirements are shown in Table 2 for four operation modes.

A variety of shapes were considered: a pillbox shape, a toroidal shape previously designed for the Next Linear Collider (NLC) at LBNL [10], and an elliptical shape. The geometry chosen for the 704 MHz cavity is a variation of the toroidal shape. It is 213.1 mm long, has 165.0 mm radius at the equator with rounded corners (rounding radius 86.1 mm), and an 82 mm beam pipe diameter, as shown in Figure 2(G). It differs from the NLC shape [10] by addition of a 40.9 mm straight section at the equator to ease fabrication of the FPC, tuner, vacuum pump and RF pickup ports. The vacuum pump port, shown in Figure 2(F), was designed 90 degrees away azimuthally from the FPC port (Figure 2(C)). A metal mesh is installed on the port flange to block RF field from penetrating toward the vacuum pump. Two RF pickup ports are located on the side opposite to the vacuum pump port. A hook-shaped coupler is used on each port to get a maximum of 2 W power out of the cavity at 250 kV.

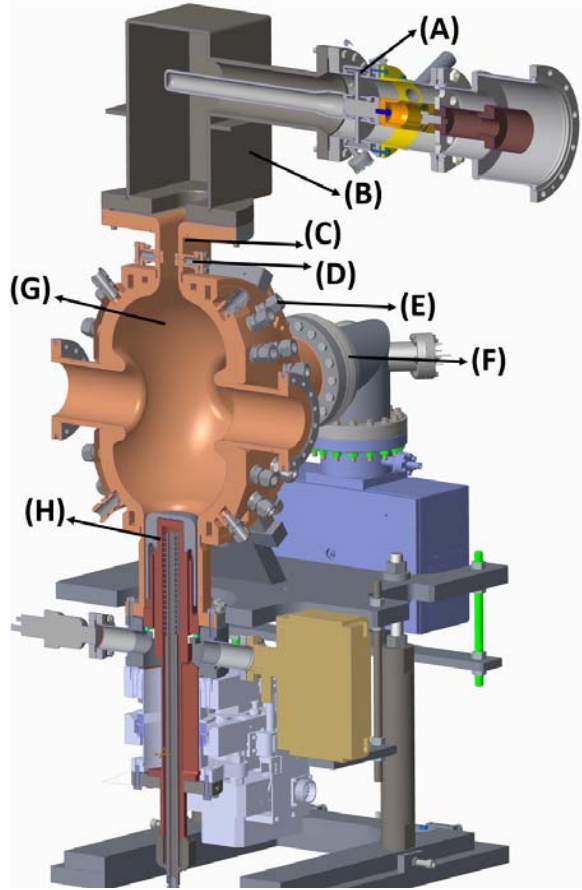

Figure 2. 704 MHz cavity cross section view. (A) Toshiba RF window; (B) WR1150 waveguide to coaxial transition piece; (C) FPC port; (D) FPC tuner; (E) cavity water cooling channel; (F) vacuum pump; (G) cavity body; (H) main frequency tuner.

## III. HOM consideration

As described in [2], 32 gaussian laser pulses with 0.6 mm rms length are stacked together with a 0.75 mm interval to form a 24 mm "flat-top" pulse. It is used to drive a DC photoemission gun [11] with a multi-alkali ($CsK_2Sb$ or $NaK_2Sb$) cathode.

Table 2. Requirements to the 704 MHz cavity at different operation modes.

| Operation mode | 1.6 MeV | 1.6 MeV CW | 2.0 MeV | 2.6 MeV |
|---|---|---|---|---|
| Bunch charge [pC] | 130 | 120.8 | 170 | 200 |
| Value of 9 MHz* [MHz] | 9.104 | 9.104 | 9.187 | 9.256 |
| Bunches per macrobunch (9 MHz) | 30 | CW | 30 | 24-30 |
| Beam Current [mA] | 35.9 | 85.0 | 47.0 | 44.2-55.3 |
| $h$ of 704 MHz | 9279 | 9279 | 9195 | 9127 |
| Voltage [kV] | 57.0 | 57.0 | 77.8 | 250.7 |
| Phase [degree] | -78.9 | -78.9 | -76.2 | -15.4 |
| Acceleration [keV] | 11.0 | 11.0 | 18.6 | 241.7 |
| Power to beam [W] | 394.1 | 933.2 | 871.5 | 13354 |
| Power on cavity [W] | 492.3 | 492.3 | 917.1 | 9523 |
| FPC coupling coefficient | 1.80 | 2.90 | 1.95 | 2.40 |

*In this paper 9 MHz means the harmonic number $h$ at 120 of RHIC revolution frequency.

To calculate the bunch-to-bunch energy spread from the long-range wake field of the longitudinal modes, a straightforward way is to "shift and stack" (based on the bunch pattern) the short-range wake field of a single bunch, which can be simulated using CST particle studio$^{TM}$, to get the final result.[2] This method is not used not only because it is time consuming (need to calculate the single bunch wake field until all modes are damped, and the "shift and stack" is also time consuming), but also because one cannot apply the "worst case scenario", which is introduced in the next paragraph, by shifting the HOM frequency around, since the actual HOM frequency in the cavity might deviate from the RF simulation due to fabrication errors, thermal and mechanical deformations, frequency tuning of the fundamental mode, etc.

We have employed a different method. First, an eigenmode simulation is done using CST Microwave Studio$^{TM}$, with the simulation frequency ranged from the fundamental mode to the first longitudinal cut-off, $f_c$, of the beam pipe. The single-bunch wake potential is then constructed using the eigenmode simulation results and is compared with the CST Particle Studio$^{TM}$ result. The multi-bunch wake potential is calculated by using the "shift and stack" method on the single bunch wake potential, similar to the calculation of "flat-top" short range wake field. The eigenmode simulation results are then treated with "worst case scenario" by artificially changing the resonance frequency of each HOM to a multiple of 9 MHz, and for those modes that are close (±20 MHz) to the multiples of 704 MHz.

For the single-bunch wake field reconstructed from the eigenmode results, bunches with three different shapes were investigated: a delta function (point) bunch, a rectangular (flat-top) bunch, and a gaussian bunch. All wake potentials are normalized to a bunch charge of one coulomb. For narrow band resonators with $Q \gg 1$, the wake potential from each longitudinal mode can be calculated as:

$$W_z(\tau) = \frac{\omega R}{2Q} e^{-\frac{\omega \tau}{2Q}} \cos(\omega \tau) H(\tau)$$

for the delta function bunch [12], where $H$ is the Heaviside step function;

$$W_z(\tau) = \begin{cases} \dfrac{1}{T}\dfrac{R}{4Q^2} e^{-\frac{\omega\tau}{2Q}}\left[ e^{\frac{\omega\tau}{2Q}} - \cos(\omega\tau) + 2Q\sin(\omega\tau) \right] \\ \qquad\qquad \text{for } 0 \le \tau \le T; \\ \\ \dfrac{1}{T}\dfrac{R}{4Q^2} e^{-\frac{\omega\tau}{2Q}} \{ e^{\frac{\omega T}{2Q}} \cos[\omega(T-\tau)] + \\ \qquad 2Q e^{\frac{\omega T}{2Q}} \sin[\omega(T-\tau)] - \\ \qquad \cos(\omega\tau) + 2Q\sin(\omega\tau) \} \quad \text{for } \tau > T \end{cases}$$

for the rectangular bunch; and

$$W_z(\tau) = \frac{\omega_r R_s}{4 Q_r} e^{-\frac{\tau^2}{2\sigma^2}} Re\left\{ \left(\frac{j}{2Q_r} + 1\right) w\left(\frac{z_1}{\sqrt{2}}\right) \right\}.$$

for the gaussian bunch [12] with

$$z_1 = \omega_r \sigma + j\left[\frac{\omega_r \sigma}{2Q_r} - \frac{\tau}{\sigma}\right]$$

and $w$-function related to the error function of complex argument:

$$\int_0^{+\infty} dt\, e^{-a^2 t^2 + jzt} = \frac{\sqrt{\pi}}{2a} w\left(\frac{z}{2a}\right).$$

In the above equations, $\omega$, $R$, $Q$ are the angular resonance frequency, shunt impedance and quality factor of a mode, $T$ is the bunch duration, and $\sigma$ is the rms bunch length for the gaussian distribution. Simulations showed that with the above-mentioned three different bunch shapes the difference between wake potential is within 20%. The delta-function bunch was selected for further simulation since it gives the highest values.

The single-bunch wake potential is then reconstructed by summing up the wake potential from all longitudinal HOMs up to $f_c$. The result for a 1 cm rms long single gaussian bunch is shown in Figure 3, with the top plot for the 704 MHz cavity and the bottom plot for the 2.1 GHz cavity, both are with 1 m wake length. The blue dotted curves are the reconstruction from eigenmode simulation, and red solid curves are the simulation results from CST Particle Studio$^{TM}$. Please note the "worst case scenario" is not applied here for fair comparison. One can see slight deviation between two curves as the reconstruction took into account only HOMs below $f_c$. Since HOMs higher than that typically leak out the beam pipe, they have low $Q$ and will have limited effect on bunch-to-bunch energy spread caused by long-range wake potential. Different boundary conditions applied on the beampipe ports (electric short for Eigenmode, and absorbing boundary for Particle Studio) that cause slightly different $Q$ values also contribute to this difference. A maximum of 20% error on the reconstruction of wake potential is estimated.

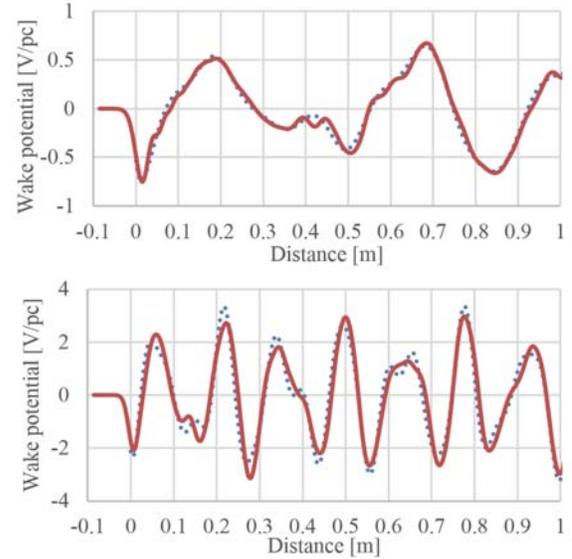

Figure 3. Single-bunch wake potential of the 704 MHz (top) and 2.1 GHz (bottom) cavities for a 1 cm rms long gaussian bunch. Blue dotted curves are the reconstruction from eigenmode simulation, and red solid curves are the simulation results from CST Particle Studio$^{TM}$.

With the reconstructed single-bunch wake potential for the "worst case scenario", we then used the "shift and stack" method to get the multi-bunch multi-train wake potential, with 30 continuous bunches at the 704 MHz repetition frequency forming bunch trains at the 9 MHz repetition rate with ~40% duty factor. The results are consistent with the method proposed by Kim et al. [13] since both methods are based on the delta function structure.

### A. 2.1 GHz Cavity

For the 2.1 GHz cavity, adjusting the end cell length might increase $R_0$. However, it will also increase the $R/Q$s of the same order modes (SOMs) thus increasing the energy spread. For example, with the end cell lengthened by 10 mm, for $TM_{010}$ $\pi$ mode, the $R_0/Q$ will decrease by 2.7% and the $Q$ will increase by 7.4%, which gives a 5% increase in $R_0$. However, alongside this change, the $R/Q$ of the 2.099 GHz SOM increases from 0.07 $\Omega$ to 9.72 $\Omega$. The frequency of this mode is mainly affected by the drift tube diameter between the cells, which determines the cell-to-cell coupling strength. It is not going to be a multiple of 704 MHz since it is 13 MHz away from 2.112 GHz. However, it is possible for its frequency to be a multiple of 9 MHz, if the cell-to-cell coupling strength is altered, thus producing a high voltage fluctuation, estimated to be ~4 kV, corresponding to the momentum spread of $\pm 2.0 \cdot 10^{-3}$, outside the required range. If the beam pipe diameter is set at 47.6 mm, the same as the diameter of the inter-cell drift tubes, the HOMs at 4.717 GHz and 4.771 GHz will give a 0.57 kV voltage fluctuation, corresponding to $\pm 2.9 \cdot 10^{-4}$ $dp/p$. Setting the beam pipe diameter at 48.9 mm allows these HOMs to propagate out of the cavity, so that both $R/Q$s and $Q$s are suppressed. Therefore we chose the length of the vacuum portion of the end cell to be ($\lambda/2$ - 10) mm, and the beam pipe diameter to be 48.9 mm to reduce the SOM/HOM induced energy spread in this cavity [3]. With the 48.9 mm diameter beam pipe, the cut-off frequency is 3.60 GHz for $TE_{11}$ mode and 4.70 GHz for $TM_{01}$ mode.

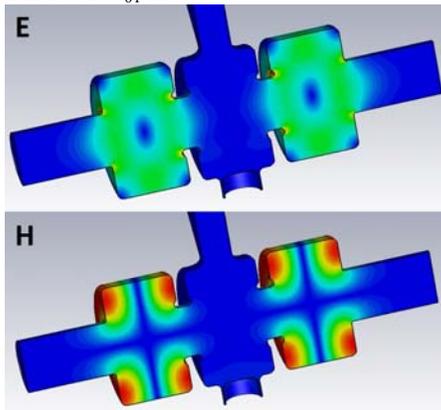

Figure 4. Field pattern of the important 3.2808 GHz HOM in the 2.1 GHz warm cavity: a) $E$ field in the cross section; b) $H$ field in the cross section.

After taking into account these considerations, there is only one mode, at 3.2808 GHz, that could drive the energy spread out of range if it beats with 9 MHz. It is the $TM_{011}$ $2\pi/3$ mode, with $R/Q$ of 73.8 $\Omega$ and $Q$ of $1.76 \cdot 10^4$. There are two SOMs associated with this mode, at 3.2574 GHz with $R/Q$ of 10.5 $\Omega$, and at 3.3034 GHz with $R/Q$ of 34.6 $\Omega$. Both modes have $Q$ factors about the same as the $2\pi/3$ mode, and produce sub-kV range voltage fluctuation, which is still within the requirement. The field pattern of the 3.2808 GHz dangerous HOM is shown in Figure 4. With this HOM, a bunch pattern with 31 bunches per train gives higher voltage fluctuation than that with 30 bunches per train. The reason is that its frequency is close to 4.66 times of 704 MHz, the repetition rate of the 30 or 31 bunches, so that the voltage fluctuation caused by this mode gets cancelled every three bunches, thus 30 bunches produce a relatively low voltage, and with 31 bunches, the last bunch produces a relatively large, $\pm 1.57$ kV, voltage fluctuation, corresponding to a $\pm 7.9 \cdot 10^{-4}$ $dp/p$ for a 2 MeV beam. If we shift the 3.2808 GHz mode 0.5 MHz away from the harmonic of 9 MHz, it gives voltage fluctuation of 0.57 kV, corresponding to $\pm 2.9 \cdot 10^{-4}$ $dp/p$.

Similar to [2], this 0.5 MHz difference can be reached by carefully choosing the electron beam kinetic energy, shown in Figure 5 (top plot). Any kinetic energy, for which the plot is above the dotted line, can guarantee that $TM_{011}$ $2\pi/3$ mode is 0.5 MHz away from the harmonic of 9 MHz.

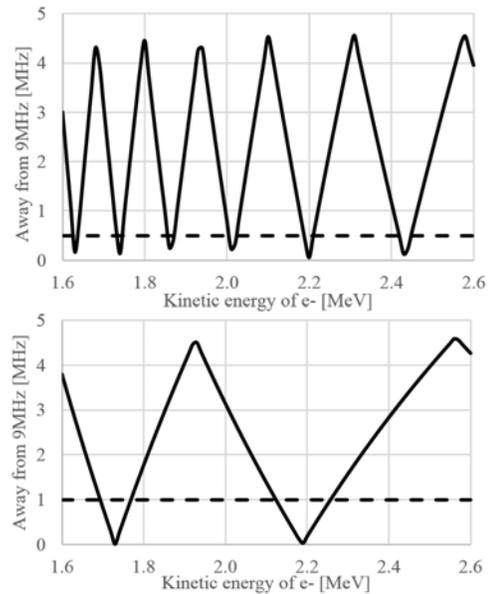

Figure 5. The choice of the electron beam kinetic energy to push the harmonic of 9 MHz away from the dangerous HOM, with the top plot for the 2.1 GHz cavity and the bottom plot for the 704 MHz cavity.

Solid lines show the frequency difference between the HOM frequency and the closest harmonic of 9 MHz for different electron beam kinetic energies. Dotted lines show the 0.5 and 1.0 MHz dangerous zones for the 2.1 GHz and 704 MHz cavities, respectively. The electron beam kinetic energy should be chosen so that the solid lines is above the dotted lines.

Eight fixed tuners, shown in Figure 1(M), are evenly distributed along the equator of the end cells, with four on each cell, 90 degrees apart. These fixed tuners are used to tune the fundamental mode frequency towards 2.112 GHz (in combination with the main tuner), and to tune the 3.2808 GHz HOM (main tuner will not affect this mode because the field in the center cell is small) away from the multiples of 9 MHz. They are flush with the cavity inner surface in the nominal position. With one fixed tuner inserted 5 mm into the cavity, the $TM_{010}$ $\pi$ mode frequency increases by 0.36 MHz, and the frequency of the 3.2808 GHz HOM increases by 2.86 MHz. This tuning was done before evacuating the cavity. Jam nuts are used to ensure that the tuners stay in position during operation.

### B. 704 MHz Cavity

If beam pipe diameter for the 2.1 GHz was used for 704 MHz cavity, monopole HOMs at 1.087, 2.818, 3.523 and 4.227 GHz would have been trapped and the HOM-induced bunch-to-bunch voltage fluctuation could be up to ~7 kV. Increasing the beam pipe to an 82 mm diameter, with the cut-off frequency of 2.14 GHz for $TE_{11}$ mode and 2.80 GHz for $TM_{01}$ mode, only the 1.087 GHz mode remains trapped. The drawback is that the fundamental mode R/Q degrades, increasing power dissipation in the cavity by ~25%.

The 31 bunch configuration gives us a 0.86 kV voltage fluctuation and the 30 bunch configuration gives us 2.04 kV. With the 1.087 GHz mode 1.0 MHz away from the harmonic of 9 MHz, the voltage fluctuation changes to 0.43 kV for 31 bunch configuration and 0.62 kV for 30 bunch configuration, corresponding to a maximum $\pm 3.1 \cdot 10^{-4}$ dp/p for a 2 MeV beam.

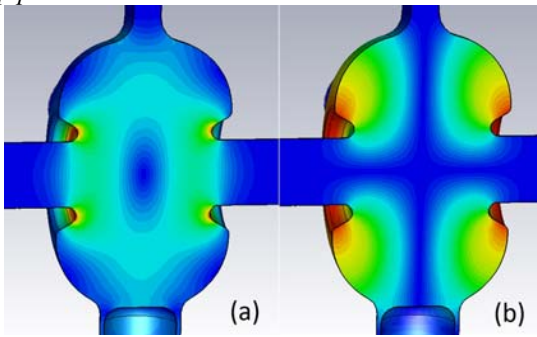

Figure 6. Field pattern of the important 1.087 GHz HOM in the 704 MHz warm cavity: a) E field in the cross section; b) H field in the cross section.

The field pattern of the 1.087 GHz HOM is shown in Figure 6. It is a $TM_{011}$ mode. Figure 5 (bottom plot) shows the choice of electron beam kinetic energy. Since this is a single-cell cavity, one cannot tune the fundamental mode and this specific HOM independently, thus no fixed tuner like in the 2.1 GHz cavity was designed.

## IV. Frequency tuner design

For the 2.1 GHz cavity, the main frequency tuner is designed to be a folded coaxial structure, shown in Figure 1(N), with a zoom-in plot in Figure 7. The cavity fundamental mode couples to a $TE_{11}$-like mode in the coaxial line. This mode has a much higher cut-off frequency than that of the fundamental $TM_{010}$ $\pi$ mode, and it does not easily transforms into the TEM mode.

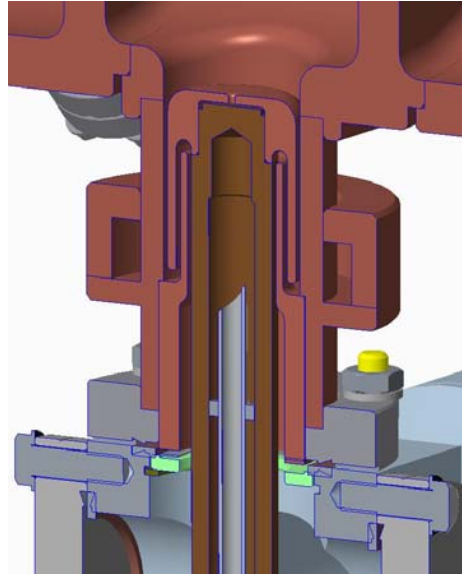

Figure 7. Cross section of the folded coaxial tuner.

### A. 2.1 GHz Cavity

For simplicity, we used a straight (non-folded) coaxial tuner model to estimate the length of the structure. The power leakage from the coaxial section is calculated by treating the end of the tuner as a waveguide port and calculating the power of the first 5 modes leaking out of the port. The results showed that a structure longer than 140 mm is needed to attenuate the power by -40 dB. Thus, in the case of a folded design, the tuner should be roughly 50 mm long. Additional length is needed to integrate bellows with the Lesker™ motor with part number 23HT18C230L500. It is important to note that a precise alignment of the tuner is necessary in this design. If the top of the tuner is shifted off axis or tilted by 0.1 mm, a factor of 40 times RF power is going to propagate out of the coaxial structure.

This design was simulated for tuner penetrations from -6 mm to +6 mm, for which the fundamental frequency changes from -1.2 MHz to +2.9 MHz with respect to the nominal. Simulations showed that multipacting barriers exist in this coaxial structure starting at 70 kV accelerating voltage. The tuner cap is designed to be demountable to provide the option of TiN coating for multipacting suppression [14].

### B. 704 MHz Cavity

The 704 MHz tuner is a 3:1 scaled up version of the 2.1 GHz tuner. It is located opposite to the FPC port, shown in Figure 2(H). With the tuner penetration varying from -12 mm to +12 mm, the frequency changes from -1.0 MHz to +1.7 MHz with respect to the nominal frequency. From simulation, multipacting is expected to start at 38 kV.

## V. FPC design

### A. 2.1 GHz Cavity

For the 2.1 GHz cavity, the FPC is critically coupled to the cavity without beam. In this case there is no strong standing wave along the FPC waveguide, and position of the FPC RF window is not critical with respect to the standing wave pattern. The RF window should not be in the line of sight of the electron beam to avoid charging of its ceramics, which may result in arcing through and creation of pinhole vacuum leaks [15].

A JLab C100/C50 RF window, shown in Figure 1(C), with a zoom-in view in Figure 8 (left), was adopted for this cavity. It is placed after a 90-degree E-bend of the 298 mm long JLab530 rectangular waveguide with dimensions 134.4 mm by 25.0 mm, shown in Figure 1(F). $TE_{10}$ cutoff frequency of the JLab530 waveguide is 1.134 GHz and $TE_{20}$ cutoff frequency is 2.266 GHz. The cavity side of the RF window is under vacuum, with a pumping port and a view port close to the window, Figure 1(D) and (E). The amplifier side is in air.

This RF window was originally designed for 1.5 GHz application at JLab, two knob tuners, shown in Figure 1(B), with a zoom-in view in Figure 8 (left), are placed on the air side to optimize the match at 2.112 GHz. The waveguide is then tapered to WR430 with dimensions 109.2 mm by 54.6 mm, shown in Figure 1(A). The notch with less than -20 dB reflection at ~2.1 GHz in $S_{22}$ is sensitive to the position of two knobs. With the knobs shifting by 0.2 mm along the waveguide, the notch frequency shifts by 26 MHz. The knob shifting is realized by rotating a bolt with a eccentric knob tip, shown in Figure 8 (left). Multipacting simulations found no multipacting on the RF window in the accelerating voltage range from 1 kV to 250 kV with 1.25 kV step size.

The FPC is designed to be critically coupled to the cavity with ideal $Q$. Surface roughness, brazing joints, as well as other imperfections, might give a degradation on $Q$ value, estimated to be up to 30%. Two FPC tuners, shown in Figure 1(H) and Figure 8 (right), are designed on the neck of the FPC port to adjust the coupling so that critical coupling can be met with different $Q$ value. Two 12.7 mm diameter knobs with 5 mm insertion can adjust the $Q_{ext}$ from 20,000 to 12,800.

The 2.1 GHz cavity operates at a beam phase of 180 degrees, decelerating the beam. In this case, when the beam current increases, the cavity gets more power from the beam, thus needing less power from an RF amplifier. Figure 9 shows the calculation results of forward and reflected power at the FPC versus the beam current. The RF amplifier should provide 9 kW power to the cavity, and the reflected power will be less than 4 kW with beam current up to 50 mA. Considering the waveguide loss, the amplifier should have output power around 14 kW. The beam current, shown in Table 2 is for both cavities, has four discrete levels: 35.9, 47.0, 55.3 and 85.0 mA. A maximum of 2.2 kW, and a minimum of 330 W, forward power is needed during the operation, corresponding to 35.9 and 85.0 mA, respectively. The 14-kW amplifier has good linearity at low power, so the cavity can be controlled by the RF system while the beam current is ramping up to 85.0 mA.

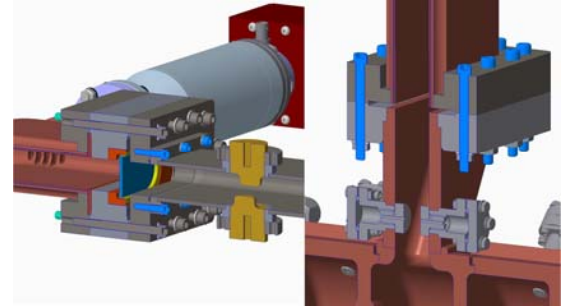

Figure 8. Zoom-in view of the JLab RF window with pumping port and eccentric knob (left, shown in yellow) and FPC coupling slot on the cavity with FPC tuners (right, shown in grey).

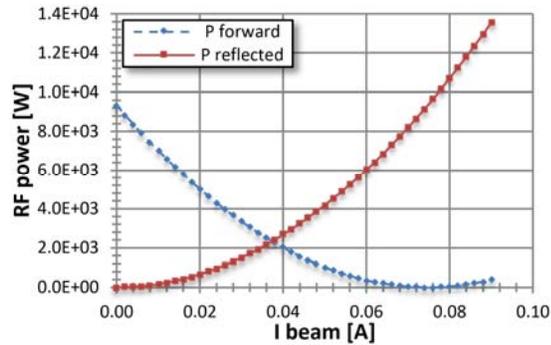

Figure 9. A calculation of forward and reflected power at the 2.1 GHz cavity FPC versus the beam current at 250 kV cavity voltage.

### B. 704 MHz Cavity

A Toshiba SNS-type coaxial window [16], shown in Figure 2(A), is used on the 704 MHz cavity. The window is connected to the cavity using a WR1150 waveguide to coaxial line vacuum transition, as shown in Figure 2(B). Multipacting simulations of this structure were performed from 10 kV to 500 kV accelerating voltage with 10 kV step size, and no multipacting was found.

To set the cavity critically coupled during operation with beam, the FPC coupling coefficient would have to change from 1.8 to 2.4, listed in Table 2. As the FPC is not adjustable, the coupling coefficient is set to 2.4 providing unity coupling at the highest required power level. Similar to the 2.1 GHz cavity, two FPC tuners, shown in Figure 2(D), are designed so that this requirement can be met.

## VI. Thermal and Mechanical Simulations

3-D coupled RF-thermal-structural analysis of the cavities has been performed using ANSYS$^{TM}$ Multiphysics finite element code to confirm the structural stability and to minimize the frequency shift resulting from heating and structural expansion.

### A. 2.1 GHz Cavity

The 2.1 GHz cavity, with a 9.5 kW power dissipation at 250 kV accelerating voltage, is cooled using 26°C water at 4 GPM flow rate. Cavity cooling channels are shown in Figure 1(K). The maximum temperature is 69°C on the nose cones, shown in Figure 10 (left). The tuner is cooled with 26°C water at 3 GPM flow rate. Tuner cooling channel is shown in Figure 1(Q). The maximum temperature is 155°C, as reported in [17], and shown in Figure 10 (right). At 250 kV, mechanical simulation showed a maximum radial deformation at 0.04 mm and a maximum axial deformation at 0.13 mm, with a maximum thermal stress near the nose cones at $4.8 \cdot 10^7$ Pa. A frequency shift of -1.2 MHz is associated with these deformations. The design is further optimized to compensate this effect by fine tuning the cavity diameter to shift the cavity resonance frequency up by 1.2 MHz at room temperature without high RF power.

### B. 704 MHz Cavity

The 704 MHz cavity was originally designed for an accelerating voltage of 430 kV, resulting in 35.5 kW power dissipation in the cavity walls. In the final LEReC configuration, the maximum operating voltage will be only 250 kV with 9.5 kW wall dissipation power. Thermal and mechanical simulation were performed for the 430 kV case. The cavity is cooled using 32°C water at 20 GPM flow rate through cooling channels in the cavity walls as shown in Figure 2(E). The maximum temperature is 64°C at the blending radius of the FPC port. Mechanical simulation showed a maximum deformation at 0.14 mm, and the corresponding von Mises stress of $3.4 \cdot 10^7$ Pa. Similar to 2.1 GHz cavity, a frequency shift of -0.14 MHz is associated with these deformations and is compensated by fine tuning the cavity diameter.

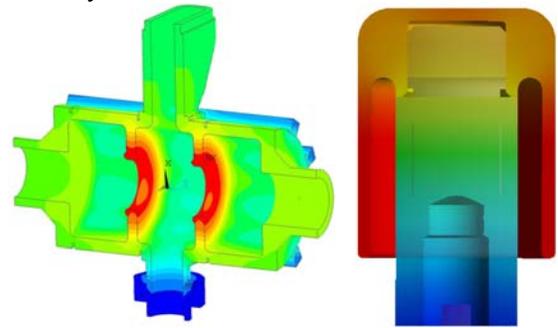

Figure 10. Temperature distribution with 9 kW total power dissipation at 250 kV accelerating voltage (left) on the 2.1 GHz cavity wall with maximum temperature at 69°C, and (right) on the tuner with maximum temperature at 155°C.

## VII. RF test

### A. 2.1 GHz Cavity

With the main frequency tuner travelling from −6 mm to +6 mm, for the 2.1 GHz cavity without FPC tuner, the FPC coupling coefficient was measured to be between 0.79 and 1.12, the frequency between 2.1112 and 2.1153 GHz. The $R/Q$ varies from 347.2 Ω to 350.7 Ω at the speed of light, and the unloaded $Q$ between 18,900 and 17,800. The reason for the FPC's external $Q$ variatrion is due to the field flatness change: with the tuner inserted deeper into the cavity, the field is pushed to the end cells, which makes the coupling weaker and thus gives a higher $Q_{ext}$. Based on the measured $Q$ value, the surface resistance showed 10% degradation from ideal polished copper.

The dangerous HOM is at 3.273 GHz, with $Q$ of 6,650 when all fixed tuners are flush with the cavity inner surface (the nominal position). With fixed tuners inserted into to the cavity, the FPC coupling coefficient drops, the $TM_{010}$ π-mode frequency increases by 0.38 MHz, and the dangerous HOM frequency drops by 2.1 MHz, for a 5 mm insertion. The fixed tuners turned out to be at the nominal position with the consideration of the dangerous HOMs in these two normal conducting cavities and in the 704 MHz SRF booster cavity [2].

With the copper main frequency tuner (without TiN coating), the cavity was high-power conditioned to 240 kV in CW mode, and 250 kV in 50% duty factor pulsed mode. The test was limited by thermal runaway of a power combiner and circulator of the RF power amplifier.

### B. 704 MHz Cavity

The 704 MHz cavity frequency was measured to be between 703.03 and 705.74 MHz, and the unloaded $Q$ between 33,700 and 31,000 with the main tuner travelling from −12 mm to +12 mm. Based on the $Q$ value, the surface resistance showed 5% degradation from ideal polished copper. With two 6.4 mm thick, 19.1 mm diameter tuners, the FPC coupling coefficient is measured to be 2.43. This cavity was high power conditioned to 250 kV in CW mode. Cavity started to outgas at 37 kV, and at around 130 kV some multipacting activities appeared at the tuner port, accompanied by a pressure spike above the vacuum trip limit set at $5·10^{-6}$ Torr ($6.7·10^{-4}$ Pa). This multipacting was conditioned away within half an hour.

## VIII. Conclusions

Two normal conducting cavities with fundamental power couplers, RF windows and frequency tuners were designed and built for the LEReC project. One cavity operates at the electron bunch repetition frequency of 704 MHz. The second cavity operates at the third harmonic. These cavities will be used for beam energy spread correction. The SOMs/HOMs in these cavities were carefully evaluated to ensure the small energy spread specification can be met. Mutiphysics simulations were performed to ensure the thermal and mechanical stability during operation. Both cavities were RF tested and high power conditioned up to 250 kV accelerating voltage to meet the operation requirements.

## ACKNOWLEDGEMENT

The work is supported by by Brookhaven Science Associates, LLC under contract No. DE-AC02-98CH10886 with the US DOE. This research used the resources of the National Energy Research Scientific Computing Center (NERSC), which is supported by the US DOE under contract No. DE-AC02-05CH11231. The authors would like to thank M. Blaskiewicz, Wencan Xu, S. Polizzo, A. Fedotov, J. Tuozzolo, N. Laloudakis and J. Butler for their help during the design and test of these two cavities.